*Sequence Analysis*

# ItLnc-BXE: a Bagging-XGBoost-ensemble method with multiple features for identification of plant lncRNAs

Guangyan Zhang[1,2,†], Ziru Liu[1,†], Jichen Dai[3], Zilan Yu[1], Shuai Liu[1], and Wen Zhang[1,*]

[1]College of Informatics, Huazhong Agricultural University, Wuhan 430070, China, [2]College of Public Management, Huazhong Agricultural University, Wuhan 430070, China, [3]Schaefer School of Engineering & Science, Stevens Institute of Technology, Hoboken, NJ 07030, United States

*To whom correspondence should be addressed.

[†]These authors contribute equally.

**Abstract**
**Motivation:** Since long non-coding RNAs (lncRNAs) have involved in a wide range of functions in cellular and developmental processes, an increasing number of methods have been proposed for distinguishing lncRNAs from coding RNAs. However, most of the existing methods are designed for lncRNAs in animal systems, and only a few methods focus on the plant lncRNA identification. Different from lncRNAs in animal systems, plant lncRNAs have distinct characteristics. It is desirable to develop a computational method for accurate and robust identification of plant lncRNAs.
**Results:** Herein, we present a plant lncRNA identification method ItLnc-BXE, which utilizes multiple features and the ensemble learning strategy. First, a diversity of lncRNA features is collected and filtered by feature selection to represent RNA transcripts. Then, several base learners are trained and further combined into a single meta-learner by ensemble learning, and thus an ItLnc-BXE model is constructed. ItLnc-BXE models are evaluated on datasets of six plant species, the results show that ItLnc-BXE outperforms other state-of-the-art plant lncRNA identification methods, achieving better and robust performances (AUC>95.91%). We also perform some experiments about cross-species lncRNA identification, and the results indicate that dicots-based and monocots-based models can be used to accurately identify lncRNAs in lower plant species, such as mosses and algae.
**Availability:** source codes are available at https://github.com/BioMedicalBigDataMiningLab/ItLnc-BXE.
**Contact:** zhangwen@mail.hzau.edu.cn (or) zhangwen@whu.edu.cn
**Supplementary information:** Supplementary data are available at *Bioinformatics* online.

## 1 Introduction

The recent improvements in high-throughput sequencing have led to the identification of numerous novel gene sequences (Chalmel, et al., 2014; Matera, et al., 2007). As a consequence, the source of coding and non-coding RNAs has been greatly enlarged. Long non-coding RNAs (lncRNAs) are a class of RNA molecules that not encode proteins, with lengths exceeding 200 nucleotides (Liu, et al., 2015). Although lncRNAs were thought to be transcriptional noise at first, increasing works demonstrate that they exert significant impacts on many biological processes, such as tissue development and external stimuli response (Chekanova, 2015; Kim and Sung, 2012; Zhang, et al., 2013).

Since only a few lncRNAs have been annotated, many machine learning-based methods have been proposed for lncRNA identification, such as CPC2 (Kang, et al., 2017), CPAT (Wang, et al., 2013), PLEK(Li, et al., 2014) and etc. CPC2 employed an SVM model using RBF kernel to distinguish coding RNAs from non-coding RNAs. CPAT used the logistic regression (LR) for novel lncRNA identification. PLEK applied a computational pipeline based on an improved k-mer scheme and an SVM algorithm. These methods were all alignment-free, which implied that they only made use of features derived directly from sequences. For example, CPC2 constructed a feature set composed of four intrinsic features, which were peptide length, isoelectric point, Fickett TESTCODE score and open reading frame (ORF) integrity, while CPAT adopted ORF length, ORF coverage, Fickett TESTCODE score and hexamer score.

In the history of lncRNA identification, the focus has always been human and animals, but few methods can be used for the plants. Plant lncRNAs are different from animal lncRNAs and may have distinct characteristics. Most of the plant lncRNAs regulate gene expression through multiple mechanisms, such as target mimicry, transcription interference, histone methylation and DNA methylation, and play essential roles in flowering, male sterility, nutrition metabolism, biotic and abiotic stress and other biological processes as regulators in plants (Liu, et al., 2015). The insufficiency of lncRNAs remains one of the major problems in plants, and most popular databases have a preference for collecting animals lncRNAs. With increasing demands in plant lncRNAs, several databases, such as RNAcentral, Ensembl Plants and CANTATAdb, began to collect plant lncRNAs. Still, there are many plant lncRNAs remain to be annotated. Therefore, it is desirable to develop a computational method for accurate identification of plant lncRNAs.

As far as we know, two methods have been proposed for plant lncRNA identification: PLncPRO (Singh, et al., 2017) and PLIT (Deshpande, et al., 2019). PLncPRO used some software, such as BLASTX to extract features. Based on a total of 71-dimensional features, PLncPRO then employed the random forest algorithm for RNA identification. Using PLncPRO models, they discovered some high-coincidence lncRNAs in rice and chickpea under abiotic stress conditions. In PLIT, seven ORF and sequence-based features, and six codon bias features were extracted from training data. PLIT adopted a feature selection process that combined the Least Absolute Shrinkage and Selection Operator (LASSO) with iterative Random Forests (iRF) to identify a list of optimal features. After that, a random forest classifier was used for plant lncRNA identification. More comprehensive studies are in demand for the plant lncRNA identification.

In this work, we present a plant lncRNA identification method ItLnc-BXE, based on multiple features and ensemble learning strategy. We collect 23 types of features that fall into four categories, and ReliefF-GA feature selection method is adopted to determine an optimal feature subset for a specific species. The subset is used to represent lncRNAs. After that, we construct the ItLnc-BXE model. We compile n data subsets by sampling data from the training dataset and accordingly build n base learners using extreme gradient boosting (XGBoost). Base learners are then combined using LR to develop the final ItLnc-BXE model. The performance of ItLnc-BXE models are evaluated on the datasets of six plant species with different lncRNAs/pcts ratios. When compared with PLIT and PLncPRO, ItLnc-BXE produces better results, which results from three aspects: (1) multiple features provide diverse information about plant lncRNAs, (2) ReliefF-GA method reduces redundancy between features, (3) ensemble learning strategy utilizes strengths from base learners.

## 2 Methods

### 2.1 Datasets

Here, we collect plant RNA transcripts from three databases, i.e., CANTATAdb version 2.0 (Szcześniak, et al., 2015), Ensembl Plants (Bolser, et al., 2016) and RNAcentral (The RNAcentral Consortium, 2014). CANTATAdb is an authoritative and comprehensive database of computationally identified plant lncRNAs, and currently contains 239,631 lncRNAs from 39 species. All lncRNAs of six plant species (Arabidopsis thaliana, Solanum tuberosum, Oryza sativa, Hordeum vulgare, Physcomitrella patens, Chlamydomonas reinhardtii) are downloaded from CANTATAdb, and used as positive instances. The pcts of six species are downloaded from Ensembl Plants, and used as negative instances. As the pcts of Hordeum vulgare and Physcomitrella patens are much more than those of other species, we randomly select 60000 Hordeum vulgare pcts and 20000 Physcomitrella patens pcts. For the other four species all pcts in the database are downloaded. To construct reliable datasets, we take three steps to preprocess raw data:

*Step1. removing invalid sequences*

First, we remove lncRNAs and pcts that lack annotations from raw data. Second, pcts in raw data may also include non-coding RNAs, such as lncRNAs, rRNAs and tRNAs, and we remove these non-coding RNAs, according to the annotations in RNAcentral (Simopoulos, et al., 2018).

*Step2. removing redundant sequences*

CD-hit (Li, 2006) is a widely used program for clustering protein or nucleic acid sequences with high efficiency, helping remove the highly similar sequences. We use CD-hit as a filter to remove redundant lncRNAs and pcts with a similarity threshold of 80%.

*Step3. constructing datasets*

For each species, we randomly choose 10% of data as the independent dataset for feature selection. The rest, 90% of data are taken as the main dataset for cross-validation. Finally, the benchmark datasets of six species are constructed (Table 1).

**Table 1.** Benchmark plant RNA transcript datasets of six species.

| Species | Main datasets | | Independent datasets | |
|---|---|---|---|---|
| | lncRNAs | pcts | lncRNAs | pcts |
| Arabidopsis thaliana (A) | 3357 | 28944 | 372 | 372 |
| Solanum tuberosum (S) | 3926 | 37150 | 436 | 436 |
| Oryza sativa (O) | 2059 | 34224 | 228 | 228 |
| Hordeum vulgare (H) | 5260 | 29039 | 584 | 584 |
| Physcomitrella patens (P) | 1056 | 14851 | 117 | 117 |
| Chlamydomonas reinhardtii (C) | 2618 | 16733 | 290 | 290 |

For each species, we take all lncRNAs as positive instances, and randomly select pcts as negative instances with the ratios (lncRNAs/ pcts) of 1:1, 1:3 and 1:5, respectively.

### 2.2 Feature extraction

As obtaining information directly from RNA sequences is difficult, we consider to transfer each sequence into a vector of digital features. So, we collect diverse plant lncRNA features from the published scientific literature (Kang, et al., 2017; Tong and Liu, 2019; Wang, et al., 2013). All collected features are classified into four categories: sequence-based features, ORF features, codon-based features, and alignment-based features. All features are summarized (Table 2), and we will give a brief description of each feature.

**Table 2.** Summary of features in this work

| Type | Feature | Dimension | Annotation |
|---|---|---|---|
| Sequence based features | Length | 1 | used |
| | GC content | 1 | used |
| | Hexamer | 1 | used |
| | Fickett | 1 | used |
| | CTD | 30 | new |
| | PI | 1 | new |
| | GRAVY | 1 | new |
| | Instability | 1 | new |
| ORF based features | ORF | 1 | used |
| | ORF-integrity | 1 | new |
| | ORF-coverage | 1 | used |
| | FF-score | 1 | used |
| Codon based features | FOP | 1 | used |
| | CUB | 1 | used |
| | RCBS | 1 | used |
| | EW | 1 | used |
| | SCUO | 1 | used |
| | RSCU | 61 | used |
| | Trimers | 64 | used |
| Alignment based features | Numbers of hits | 1 | used |
| | Significance score | 1 | used |
| | Total bit score | 1 | used |
| | Frame entropy | 1 | used |

*Note*: Annotations 'used'/ 'new' mean that features have/ haven't been used in the plant lncRNA identification.

### 2.2.1 Sequence-based features

Sequence-based features are directly extracted from transcripts or indirectly calculated by them.

Transcript length ('Length') is one of the most fundamental features used to distinguish lncRNAs from pcts as lncRNAs' length not exceeding 200 nucleotides. GC content is the percentage of guanine (G) and cytosine (C) in four kinds of nitrogenous bases, including adenine (A) and thymine (T). The study (Singh, et al., 2017) reported that the GC content in lncRNA is less rich than that in pcts. Hexamer score ('Hexamer') (Tong and Liu, 2019) is calculated based on the occurrence of hexamer along a sequence. Fickett score ('Fickett') is a simple linguistic feature that distinguishes protein-coding from non-coding transcripts according to the combinational effect of nucleotide composition and codon usage bias. Composition, transition and distribution features ('CTD') considers the nucleotide composition (descriptor 'C'), transition (descriptor 'T') and distribution (descriptor 'D') of RNA sequences (Tong and Liu, 2019). 'C' describes the content of four nucleotides among the sequence. 'T' represents the percent frequency with the conversion of four nucleotides between adjacent positions, which means the content of AG (or GA, vice versa), AC, TG, TC and GC along a sequence. 'D' indicates five relative positions (0, 25%, 50%, 75%, 100%) among the transcripts of four nucleotides. Isoelectric point('PI') (Kang, et al., 2017) is the theoretical isoelectric point of a predicted peptide calculated by the ProtParam module in BioPython. Grand Average of Hydropathy ('GRAVY') (Kang, et al., 2017) value means the grand average of hydropathicity, a predicted peptide of which is calculated by the ProtParam module in BioPython. Instability provides an estimate of the stability of the protein in a test tube with a weight value of instability to different dipeptides.

### 2.2.2 ORF-based features

An open reading frame (ORF) is a portion of a gene's sequence that contains a sequence of bases and could potentially encode a protein. ORF features are fundamental ones to distinguish lncRNA from pcts.

ORF length ('ORF') is the maximum length of the ORF. Studies (Frith, et al., 2006; Wang, et al., 2013) revealed that protein-coding genes usually have long ORFs (>100 codons), while putative long ORFs in non-coding genes can hardly be observed. ORF integrity indicates whether the ORF begins with a start codon and ends with an in-frame stop codon (Kong, et al., 2007). ORF coverage is the ratio of ORF length to transcript length. It is reported that The score of ORF coverage is much lower in non-coding RNAs than in protein-coding RNAs (Wang, et al., 2013). ORF score, termed as FF-score, is extracted using Framefinder software (Singh, et al., 2017).

### 2.2.3 Codon-based features

Codon-based features are related to the different usage frequencies of codons that occur in the pcts as one specific amino acid usually can be translated from several synonymous codons.

Frequency of the optimal codons ('FOP') (Deshpande, et al., 2019) is the ratio of the number of optimal codons to a total number of synonymous codons. Codon Usage Bias ('CUB') (Deshpande, et al., 2019) is the index that estimates the differences of codon bias between test set sequences and reference set sequences. Strength of Relative Codon Bias ('RCBS') (Roymondal, et al., 2009) is an overall score of a gene that indicates the influence of RCB of each codon. RCB reflects the level of gene expression. Weighted sum of relative entropy ('EW') (Deshpande, et al., 2019) evaluates the degree of deviation from equal codon usage. Synonymous Codon Usage Order ('SCUO') (Deshpande, et al., 2019) is also a measure related to entropy-based codon bias. Relative Synonymous Codon Usage ('RSCU') (Sharp, et al., 1986) refers to the relationship between observed codon frequencies and the number of times codon. We also calculated frequencies of 64 trimers ('Trimers') among A, C, G and T to capture potential codon usage bias.

### 2.2.4 Alignment-based features

Alignment-based features are obtained by aligning all the sequences to curated sequences to observe the similarity between unpredicted transcripts and labeled ones. Different from the intrinsic properties of each transcript itself, alignment-based features are necessary.

BLAST is a useful tool for finding regions of similarity between nucleotide or protein sequences (Altschul, et al., 1997). The basic idea of BLAST is to align the query sequence with sequences in a database. Then it generates satisfying aligned word pairs, and each pair is called a 'hit'. We use BLAST program to assess whether lncRNAs have significant similarity to pcts in SWISS-PROT database (O'Donovan, et al., 2002). The following four features are extracted by parsing the BLAST output (Singh, et al., 2017). Number of hits is as a fundamental indicator in BLAST, and the number of hits for pcts is expected higher than that for lncRNAs. However, many sequences show random unimportant matches to a BLAST database, so the quality of the hit is considered using three more features: Significance score, Total bit score and Frame entropy. Significance score establishes an intuitive relationship between the e-value in BLAST and the quality of hits of a given sequence. Total bit score simply sums up all the bit scores which is a normalized measure evolved from raw alignment score in BLAST. Frame entropy indicates the way of the hits distributed in different reading frames.

### 2.3 Feature selection

Feature selection is a process of selecting most discriminative features from a set of features. This method can be used to identify and remove redundant features that do not contribute to or even decrease the accuracy of predictive models. There are two types of commonly used feature selection methods: filter methods and wrapper methods. However, filter methods ignore dependencies between features, whereas wrapper methods are inefficient in time cost (Deshpande, et al., 2019). Here, we adopt a novel feature selection method (Li, et al., 2009), called ReliefF-GA, by combining the ReliefF (Robnik-Šikonja and Kononenko, 2003) with the Genetic Algorithm (GA) (Goldberg, 1989). The integration of ReliefF and GA overcome the weaknesses of a single filter or wrapper method, thus lead to an effective selection scheme.

First, the ReliefF is applied to remove features that are not contributive or even counterproductive to classification. ReliefF is the extension of Relief algorithm (Kira and Rendell, 1992). It is more robust and can deal with incomplete data compared with the original Relief algorithm. Similar to Relief, the key idea of the ReliefF is to estimate the quality of features according to how well their values distinguish between sequences that are near to each other. After performing ReliefF algorithm, features with an importance score less than zero are removed because the threshold of zero implies whether this feature is contributed.

Next, we perform GA to obtain the optimal features subset for each species. GA is a heuristic optimization method inspired by natural evolution. In feature selection by GA, it starts with a set of candidate individuals called population. Each individual, also a combination of selected features in the population indicates a solution to the selection problem. With the initial population, it starts iterations to produce better approximations. In each generation, individuals in the population may undergo crossovers, mutations and then being selected according to their levels of fitness. To simplify our problems, binary encoding is adopted to represent the feature combination. A bit of '1' means the corresponding feature is selected, whereas '0' indicates not. Hence, a solution is converted into a binary string with length equal to the total number of features. We initialize the first generation based on feature candidates produced by ReliefF. For each combination in the population, the XGBoost classifier is built on the training set and tested on the test set. After that, we consider models' predictive AUC scores to represent the level of fitness of corresponding feature combinations. Eventually, GA obtains the optimal feature subset after a series of iterative computations.

### 2.4 Ensemble learner construction

In machine learning, an ensemble learner consists of several base learners, and each base learners will have its own classification strengths, resulting in stronger and more accurate predictions than individual base learners, and ensemble models have many successful applications in bioinformatics (Gong, et al., 2019; Zhang, et al., 2019; Zhang, et al., 2018). The bagging algorithm is a commonly used ensemble strategy (Dudoit and Fridlyand). It has an effective application of reducing the variance and improving the classification ability of the base learners in supervised learning. Here, we propose an ensemble model for the plant lncRNA identification using a bagging algorithm. First, we use the bootstrapping algorithm to generate multiple data subsets from the training dataset. The XGBoost is a scalable tree boosting system (Chen and Guestrin, 2016) that has superior performance in supervised learning, and we apply it to build multiple base learners based on data subsets. How to combine these base learners is critical and challenging work. Popular ways such as arithmetic mean and majority voting are usually utilized. We adopt an LR meta-learner in order to reduce the information redundancy between base learners. The LR meta-learner uses the outputs from base learners as inputs, and then produce a score indicating the probability of being a plant lncRNA.

### 2.5 Workflow of ItLnc-BXE

A workflow describes the process of ItLnc-BXE (Fig. 1). First, we construct the datasets for each species, including main datasets and independent datasets, and they have no overlap. Next, the feature selection is implemented for each species using the independent datasets, and the optimal feature subsets are determined for the model construction. After that, we perform cross-validation experiments based on the main datasets.

In each fold of cross-validation, we divide training data into two parts: 7/10 of data for training base learners and 3/10 of data for the LR learner. Using bootstrapping algorithm, $n$ data subsets are sampled from 7/10 of training data, based on which $n$ XGBoost learners are trained. These base leaners are applied to the prediction of the rest 3/10 of data, results of which are regarded as training data for the LR learner. In this way, the ItLnc-BXE model is constructed.

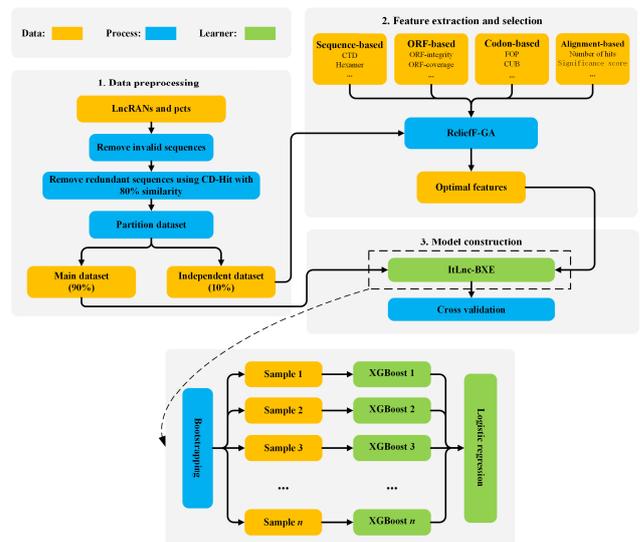

**Fig. 1.** Workflow of ItLnc-BXE that involves the following steps: construction of benchmark dataset, extraction of four categories of features and feature selection, and construction of ItLnc-BXE model. (1) We exclude wrongly annotated sequences from raw data and then use CD-Hit to remove similar sequences. Subsequently, data are divided into main datasets and independent datasets. (2) We collect 175-dimensional features from four categories and adopt ReliefF-GA to select optimal features. This process is based on independent datasets. (3) Sequences in the main datasets are transformed into feature vectors according to optimal features in step (2), on which ItLnc-BXE models are constructed. We sample training datasets into $n$ subsets, based on which $n$ XGBoost base learners are built. Then, all learners are combined using LR.

## 3 Results and discussion

### 3.1 Experimental setting

We evaluate all ItLnc-BXE models on the datasets of six species: Arabidopsis thaliana (A), Solanum tuberosum (S), Oryza sativa (O), Hordeum vulgare (H), Physcomitrella patens (P) and Chlamydomonas reinhardtii (C).

As ItLnc-BXE samples sub-datasets from training data and build several base learners on them, the number of base learners used in ensemble models ought to be determined. We perform 10-fold cross-validation to evaluate ItLnc-BXE models with different numbers of base learners using the independent datasets and determine to use five base learners for ItLnc-BXE according to the experimental results.

10-fold cross-validation is performed on the main sets to evaluate the performances of ItLnc-BXE and compared methods. We adopt popular evaluation metrics, including the area under the ROC curve (AUC), the area under the precision-recall curve (AUPR), accuracy, sensitivity (SN), specificity (SP), PRE, f1-score and Matthews correlation coefficient (MCC).

### 3.2 Feature discussion

Features are critical for distinguishing lncRNAs from pcts, and thus we consider a variety of features for the plant lncRNA identification. However, these features may make different contributions to the identification of lncRNAs from different species, and some are redundant. To make analysis, we apply the ReliefF method to score the importance of collected features, and we obtain scores of all features for each species. Then, we calculate the Pearson correlation coefficients (PCC) between scores of all features for every two species (see Table 3). PCC is used to measure the correlation of the feature ranking lists for different species. The results show that some species have relatively high correlations (>0.8) and others have comparably low correlations (<0.6), indicating that the species have the preference for features. Therefore, it is necessary to determine optimal feature subsets from candidates to build species-specific models.

**Table 3.** PCC of ReliefF importance scores between every two species.

|   | A | S | O | H | P | C |
|---|---|---|---|---|---|---|
| A | 1 | | | | | |
| S | 0.908 | 1 | | | | |
| O | 0.821 | 0.870 | 1 | | | |
| H | 0.856 | 0.808 | 0.827 | 1 | | |
| P | 0.775 | 0.834 | 0.792 | 0.667 | 1 | |
| C | 0.552 | 0.523 | 0.483 | 0.532 | 0.316 | 1 |

We adopt ReliefF-GA to select optimal feature subsets for the model construction. Since there are six species, we implement the feature selection respectively for each one using their independent datasets. Results show that ReliefF-GA greatly reduces 175-dimensional features for all species to lower dimensions ("A":89, "S":93, "O":87, "H":95, "P":90, "C":88), and refer to Supplementary Table S0 for detail. However, the optimal feature subsets are different for each species. To explore commonly used features for two species, we respectively calculated the Jaccard similarity coefficient (JSC). JSC is defined as follow:

$$J(S_1, S_2) = \frac{|S_1 \cap S_2|}{|S_1 \cup S_2|}$$

where $S_1$ and $S_2$ are two sets, $|S_1 \cap S_2|$ means the card of the intersection and $|S_1 \cup S_2|$ means the card of the union. JSC indicates the similarity between the two sets. Then, we calculate JSC for every two species (Fig. 2). The results show that JSCs range from 0.32 to 0.40, which means that these species share some features but use more different features. Further, we pay attention to those commonly used features. We find several features (Instability, one dimension from CTD, two from RSCU and three from Trimers) are shared by six species, indicating that these features are preferred in the plant lncRNA identification. Moreover, some features (Fop, Frame Entropy, three dimensions from CTD, four from Trimers and eight from RSCU) are shared by five species at most, indicating they can be commonly used in plant lncRNA identification. For more commonly used features share by more than two species, refer to Supplementary Table S1.

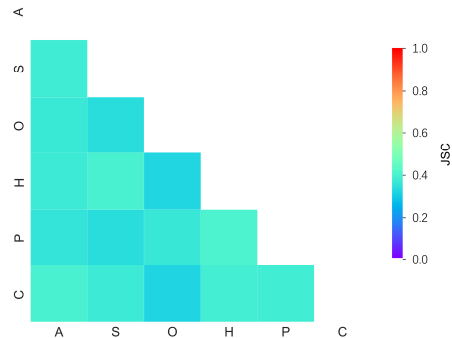

**Fig. 2.** Heatmap of JSCs between every two species. JSC ranges from 0 to 1, and the bigger JSC means two species share more features.

### 3.3 Comparison with other methods

Two machine learning methods: PLIT and PLncPRO have been presented for the plant lncRNA identification, and we adopt them for comparison. Source codes of PLIT and PLncPRO are publicly available, so we can correctly build PLIT and PLncPRO models, and then compare the performance of ItLnc-BXE with them on six different species (A, S, O, H, P, C). For each species, we consider the datasets with different lncRNAs/pcts ratios (1:1, 1:3 and 1:5). All prediction models are evaluated by using 10-fold cross-validation.

Here, we take the results on species 'A' for analysis (see in Table 4). Clearly, ItLnc-BXE produces better AUC scores than PLIT and PLncPRO in terms of the AUC scores for all species, and improvements on accuracy, SP, f1-score, MCC and PRE are also observed. Moreover, we explore how the difference between lncRNAs/pcts ratios in datasets influences the performances of prediction models. Results show that ItLnc-BXE produces similarly AUC scores and accuracy on datasets with different ratios, and the conclusion can also be drawn for the compared methods. Further, we calculate standard deviations of AUC scores of each model on these datasets, and it seems that ItLnc-BXE has lower standard deviations (0.017) than PLIT (0.273) and PLncPRO (0.071), indicating that ItLnc-BXE is robust to the data imbalance.

The results on all species are included in Supplementary Table S2-S9. In general, ItLnc-BXE produces better results than PLIT and PLncPRO on the benchmark datasets of all six species. The superiority of ItLnc-BXE is owing to several factors. First, we consider a variety of representative features that have proved to be useful in lncRNA identification. They bring enrich information for building high-accuracy models. Second, the feature selection method helps to determine the most informative features and reduce redundancy. Third, the ensemble learning strategy makes use of the strengths of base learners, thus leads to robust performances.

**Table 4.** Performance comparison of ItLnc-BXE and compared methods on species 'A'

| lncRNAs/pcts | Methods | AUC (%) | ACC (%) | AUPR (%) | SN (%) | SP (%) | MCC (%) | F1-score (%) | PRE (%) |
|---|---|---|---|---|---|---|---|---|---|
| 1:1 | ItLnc-BXE | 99.27 | 96.60 | 99.08 | 97.68 | 95.53 | 93.23 | 96.64 | 95.63 |
| | PLIT | 95.78 | 87.98 | 95.65 | 87.94 | 88.03 | 75.99 | 87.97 | 88.03 |

|  |  |  |  |  |  |  |  |  |
|---|---|---|---|---|---|---|---|---|
|  | PLncPRO | 99.09 | 95.73 | 98.85 | 98.66 | 92.79 | 91.61 | 95.85 | 93.19 |
| 1:3 | ItLnc-BXE | 99.31 | 96.60 | 97.72 | 93.57 | 97.61 | 90.95 | 93.22 | 92.88 |
|  | PLIT | 96.29 | 90.33 | 90.38 | 76.14 | 95.07 | 73.57 | 79.73 | 83.74 |
|  | PLncPRO | 99.25 | 96.66 | 97.47 | 95.12 | 97.18 | 91.23 | 93.45 | 91.83 |
| 1:5 | ItLnc-BXE | 99.30 | 97.07 | 96.25 | 90.71 | 98.34 | 89.40 | 91.15 | 91.63 |
|  | PLIT | 96.41 | 92.55 | 86.30 | 68.46 | 97.37 | 71.58 | 75.38 | 83.94 |
|  | PLncPRO | 99.23 | 96.97 | 95.92 | 91.12 | 98.14 | 89.10 | 90.92 | 90.72 |

### 3.4 Performances of cross-species identification

As discussed above, we can build species-specific ItLnc-BXE models based on datasets about species, and it is very interesting to examine the performance of ItLnc-BXE in cross-species identification. We conduct the following experiments based on the datasets of six species with ratio 1:1, and all models are built on the main sets. It's important to point out that, in cross-species identification between every two species, we test a specific model based on all data (main set and independent set) of another species. As for species self-identification, we just take the results of 10-fold cross-validation based on the main sets.

To clearly present results, we draw radar and bar figures (Fig. 3). Six plant species can be classified into four categories: dicots (Arabidopsis thaliana and Solanum tuberosum), monocots (Oryza sativa and Hordeum vulgare), the moss (Physcomitrella patens) and the alga (Chlamydomonas reinhardtii). Thus, the species in a category are visualized in a sub-figure for comparison. In general, ItLnc-BXE produces AUCs ranging from 88.60% to 99.27%. The performances of other metrics are provided Supplementary table S10-S17.

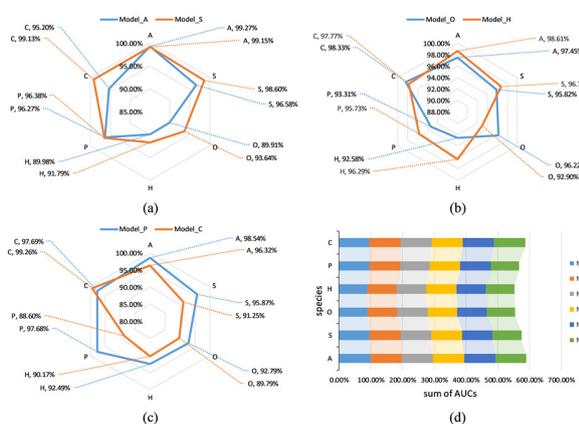

**Fig. 3.** Performance comparison (AUC scores) of ItLnc-BXE in cross-species lncRNA identification. (a) Results of dicots-based models on six species, (b) results of monocots-based model and (c) results of lower plant-based models. (d) Sum of AUC scores of six species

ItLnc-BXE models constructed on dicots produce relatively high AUCs on the moss (96.27%-96.38%) and the alga (95.20%-99.13%), but comparably low on monocots (89.91%-93.64%) (Fig. 3a). It is possible that dicots conserve plenty of biological commonness with mosses and algae in the process of evolution. More specifically, some lncRNAs in mosses and algae may have the same composition and function as those in dicots, which leads to good prediction performance of dicots-based models on the moss and the alga. Species 'A' and 'S' are both dicots, and they produce similar and very high AUCs in cross-species identification. This indicates lncRNAs in different dicot species are closely similar to each other.

Similarly, monocots models produce relatively high AUCs on both dicots (95.82%-98.61%), the moss (93.31%-95.73%) and the alga (97.77%-98.33%) (Fig. 3b). This also means lncRNAs in monocots conserve close similarity to those in dicots, mosses and algae. However, there seems to be a contradiction. As mentioned before, dicots-based models produce comparably low AUCs on monocots, indicating lncRNAs in dicots are not very similar to those in monocots. This is actually explicable, probably because lncRNAs in dicots are less abundant than those in monocots. As a lack of information on monocot lncRNAs, dicots-based models will not perform equally well on monocots. Besides, within monocots, AUCs between species 'O' and 'H' do not exceed 92.90%. We suspect that although monocots root from similar ancestors, they may gradually obtain specific biological properties, resulting in one always has defects to identify lncRNAs of other monocots.

Since the moss and the alga are both lower plants, we put them together for analysis (Fig. 3c). The moss-based model produces relatively high AUCs on dicots (95.87%-98.54%) and the alga (97.69%), but comparably low AUCs on monocots (92.49%-92.79%). This demonstrates again that lncRNAs properties in mosses are similar to those in dicots and algae, and lncRNAs in monocots are possibly far more abundant than those in mosses. The alga-based model produces comparably low AUCs on both monocots (89.79%-90.17%) and the moss (88.60%), and ranging AUCs on dicots (91.25%-96.32%). It is probably because the alga owns fewer similarities to other plants.

From another angle, we sum up the AUCs of ItLnc-BXE models for each tested species (Fig. 3d). The results show that dicots produce the highest sums (575%-589%), the moss and the alga produce relatively high sums (568%-587%), and monocots produce comparably low sums (553%-555%). The AUC sums can describe how easy lncRNAs of one species are predicted by ItLnc-BXE models based on other species. Therefore, lncRNAs of dicots, mosses and algae are more predictable than monocots; even if lack of data of lower plants, we can utilize dicots or monocots data to build models to identify lower plant lncRNAs.

## 4 Conclusion

In this work, we propose an ItLnc-BXE based on the ensemble learning and bagging algorithm to identify plant lncRNAs. ItLnc-BXE makes use of diverse features that have been proved to be useful in lncRNA identification or related works, and the feature selection method is used to select the optimal feature subset. The frame of ensemble learning further improves the performances. ItLnc-BXE constructs species-specific models on datasets of six plants respectively, and cross-validation experiments show that these models produce good performances. When compared with PLIT and PLncPRO, ItLnc-BXE yields better results on datasets of six species. ItLnc-BXE is a promising method for identifying lncRNAs from transcripts.

The studies on the features reveal that plant lncRNAs from different spices have a preference for different features but still share some features. Moreover, the studies on the cross-species lncRNA identification of six species suggest that: (1) cross-species models achieve good performances, (2) lncRNAs in dicots, mosses and algae are easy to be identified using models based on other species. Therefore, we can build ItLnc-BXE models on species with abundant data sources to identify lncRNAs in species lack of data.

## Acknowledgments

None

## Funding
This work has been supported by the National Natural Science Foundation of China (61772381, 61572368), the Fundamental Research Funds for the Central Universities (2042017kf0219, 2042018kf0249), National Key Research and Development Program (2018YFC0407904), Huazhong Agricultural University Scientific & Technological Self-innovation Foundation. The funders have no role in study design, data collection, data analysis, data interpretation, or writing of the manuscript.

*Conflict of Interest:* none declared.